\documentclass[twocolumn,tighten,times]{aastex63}
\usepackage{graphicx}
\usepackage{amsmath}
\usepackage{natbib}
\usepackage{amssymb}

\usepackage{sidecap}

\shorttitle{Fallback and late X-rays from GW170817}
\shortauthors{Metzger \& Fern\'andez}

\begin{document}

\newcommand{\be}{\begin{equation}}
\newcommand{\ee}{\end{equation}}

\title{From Neutrino- to Photon-Cooled in Three Years: Can Fallback Accretion Explain the X-ray Excess in GW170817?}

\author[0000-0002-4670-7509]{Brian D. Metzger}
\affil{Department of Physics and Columbia Astrophysics Laboratory, Columbia University, Pupin Hall, New York, NY 10027, USA}
\affil{Center for Computational Astrophysics, Flatiron Institute, 162 5th Ave, New York, NY 10010, USA} 

\author[0000-0003-4619-339X]{Rodrigo Fern\'andez}
\affil{Department of Physics, University of Alberta, Edmonton, AB T6G 2E1, Canada.}

\begin{abstract}
Excess X-ray emission from the neutron star merger GW170817 above the predicted afterglow was recently detected $t \approx 3.4$\,yr post-merger.  One possible origin is accretion onto the newly unshrouded black hole (BH) remnant.  While fall-back of bound dynamical ejecta is insufficient to generate the excess luminosity, $L_{\rm X} \sim 5\times 10^{38}$ erg s$^{-1}$, fall-back from the disk wind ejecta$-$due to their larger mass and lower velocity$-$remains a possibility.  We present hydrodynamic $\alpha$-viscosity simulations of the post-merger disk evolution which extend to timescales $t \approx 35$\,s post-merger, necessary to capture the asymptotic evolution into the radiatively inefficient regime.  Due to inefficient neutrino cooling, the BH accretion rate decays rapidly at late times ($\dot{M}_{\rm bh} \propto t^{-\beta_{\rm bh}}$, where $\beta_{\rm bh} \approx 2.4-2.8$), incompatible with the late-time excess.  However, matter falls back to the inner disk from the equatorial region more gradually, $\dot{M}_{\rm fb} \propto t^{-\beta_{\rm fb}}$ with $\beta_{\rm fb} \approx 1.43$ in our $\alpha \approx 0.03$ simulations.  By the present epoch $t \approx 3.4$\,yr, the fall-back rate has become sub-Eddington and the disk can again accrete efficiently, i.e. $\dot{M}_{\rm bh} \approx \dot{M}_{\rm fb}$, this time due to photon instead of neutrino cooling.  The predicted present-day X-ray accretion luminosity, $L_{\rm X} \approx 0.1 \dot{M}_{\rm bh}c^{2} \approx (2-70)\times 10^{38}$ erg s$^{-1}$ for $\beta_{\rm fb} \approx 1.43-1.66$, thus supports (with caveats) an accretion-powered origin for the X-ray excess in GW170817.  The suppressed BH accretion rate prior to the sub-Eddington transition, weeks to months after the merger, is key to avoid overproducing the kilonova luminosity via reprocessing.
\end{abstract}


\section{Introduction}

The neutron star merger GW170817 \citep{LIGO+17DISCOVERY} was accompanied by radiation covering the electromagnetic spectrum, from radio to gamma-rays \citep{LIGO+17CAPSTONE,Margutti&Chornock21}.  Thermal kilonova emission, powered by the radioactive decay of $r$-process elements \citep{Li&Paczynski98,Metzger+10,Barnes&Kasen13}, was observed starting 11 hours after the merger (e.g., \citealt{Coulter+17,SoaresSantos+17,Arcavi+17}) and continued to be detected in the infrared up to 74 days  \citep{Villar+18,Kasliwal+19}.  The merger was also accompanied by a broadband non-thermal synchrotron afterglow \citep{Margutti+17,Hallinan+17,Haggard+17}, created by the shock interaction of a relativistic jet interacting with the circum-merger medium (e.g., \citealt{Lazzati+18,Gottlieb+18,Granot+18,Lamb&Kobayashi18,Wu&MacFadyen19}).  Such relativistic jets are believed to give rise to short-duration gamma-ray bursts at cosmological distances, powered by the accretion of neutron star debris onto the newly formed black hole (e.g., \citealt{Narayan+92}).

After peaking at $t \approx 150$ days post-merger, the non-thermal radio and X-ray emission began to fade as a steep power-law in time, consistent with theoretical predictions of the afterglow of a jet initially directed away from our line of site \citep{Margutti+18,Nynka+18,Piro+19}.  However, a significant excess relative to the afterglow model was recently reported based on {\it Chandra X-ray Observatory} observations at $t \approx 3.4$ years \citep{Balasubramanian+21,Hajela+21,Troja+21}.  A similar excess is not seen in the radio band, requiring a spectral flattening relative to the nearly fix power-law spectral shape observed at earlier epochs.  Such a flattening is not expected in the standard afterglow model in which the spectral index of the electrons is fixed (e.g., \citealt{Hajela+21}).  Instead, its presence suggests the onset of a new component of emission which predominantly contributes with an X-ray luminosity of $L_{\rm X} \sim 5\times 10^{38}$ erg s$^{-1}$ in the $\sim$keV range, albeit with significant uncertainty due to low photon statistics and sensitivity to the method of instrumental calibration.

Two main ideas have been put forward for generating an extra late component of emission from a neutron star merger.  One possibility is that this component represents a different source of synchrotron emission, powered by the shock interaction of the kilonova ejecta with the interstellar medium (e.g., \citealt{Nakar&Piran11,Kathirgamaraju+19,Nedora+21}).  Although the bulk of the merger ejecta expanded at velocities $\approx 0.1-0.2 c$ (e.g., \citealt{Villar+17}), the dynamical ejecta can possess a small quantity of higher velocity material.  The need to generate enough high-velocity matter to explain the X-ray luminosity can in principle constrain the neutron star equation of state \citep{Nedora+21}, to which the dynamical ejecta is sensitive (e.g., \citealt{Radice+18}).  However, a drawback of this scenario is that it requires the shock-accelerated electrons to possess a flat energy distribution ($p \lesssim 2.1$, where $dN/dE \propto E^{-p}$), in tension with the larger value $p \approx 2.2$ inferred from the earlier afterglow (e.g., \citealt{Margutti+18}) and from other non-relativistic shocks such as those in supernovae (e.g., \citealt{VanDyk+94}).

Another potential source of X-ray emission occurs on much smaller scales: the accretion disk of the newly formed BH remnant\footnote{We note that a stable cooling {\it neutron star} remnant, should the equation of state permit such a merger outcome, would only generate an intrinsic X-ray luminosity $\sim 10^{35}$ erg s$^{-1}$ on a timescale of years after the merger (e.g., \citealt{Beznogov+20}) and hence would be incompatible with powering the observed X-ray excess.} being fed by the fall-back of debris from the merger \citep{Hajela+21,Ishizaki+21}.  As discussed by \citet{Hajela+21}, an accretion-powered X-ray source has several appealing features.  Firstly, the X-ray luminosity is below the Eddington luminosity, $L_{\rm Edd} \approx 8\times 10^{38}$ erg s$^{-1}$ of the $\approx 2.6M_{\odot}$ black hole remnant \citep{LIGO+17PARAMS}, consistent with the generation of a radiatively efficient accretion flow down to the innermost stable circular orbit (ISCO) of the spinning stellar-mass black hole.  Secondly, the predicted temperature of the thermal disk emission at this luminosity is $kT_{\rm eff} \sim 1$ keV, close to that of the observed excess.  Furthermore, the optical depth of X-ray photons through the kilonova ejecta can reach $\lesssim 1$ on a timescale of years after the merger (e.g., \citealt{Margalit+18}), unshrouding the black hole accretion funnel.

Perhaps the biggest open question regarding the accretion scenario is whether sufficient mass is returning to the black hole years after the merger.  Even assuming an initial binary with an unequal mass ratio, the quantity of dynamical ejecta falls short by orders of magnitude (e.g., \citealt{Rosswog07}).  On the other hand, the large mass of the kilonova ejecta from GW170817 ($M_{\rm ej} \approx 0.03-0.08M_{\odot}$; \citealt{Cowperthwaite+17,Arcavi+17,Villar+17}) excludes a dynamical origin, but is consistent with arising in outflows from the post-merger accretion disk (e.g., \citealt{Metzger+08,Metzger+09,Lee+09,Fernandez&Metzger13,Just+15,Siegel&Metzger17,Fujibayashi+18,Fernandez+19}).  If a mass comparable to $M_{\rm ej}$ were to remain gravitationally bound to the black hole and return at a sufficiently gradual rate, the observed X-ray excess could be produced \citep{Hajela+21,Ishizaki+21}.  On the other hand, if the black hole accretion rate is too large on intermediate timescales (days to months, when the ejecta is still opaque at X-ray wavelengths), this could over-produce the kilonova luminosity via X-ray reprocessing (e.g., \citealt{Kisaka+16,Matsumoto+18}).

In this Letter we address these issues by means of hydrodynamical simulations of the post-merger black hole accretion flow.  Adopting disk models 
consistent with parameters inferred from GW170817 and numerical relativity
simulations, and which are consistent with the evolution found in full 
MHD (e.g., \citealt{Fernandez+19}), we run the simulations for an unprecedently long time $\approx 35$ seconds ($\gtrsim10^{6}$ dynamical times at the black hole horizon), as necessary to follow the evolution of the marginally bound debris
into the full radiatively-inefficient stage.  Extrapolating our findings to timescales $\sim$\,yr, we demonstrate that fall-back from the disk outflow may in principle be sufficient to account for the X-ray excess of GW170817, while simultaneously evading an overproduction of the kilonova emission.

\begin{figure*}
    \centering
    \includegraphics*[width=0.8\textwidth]{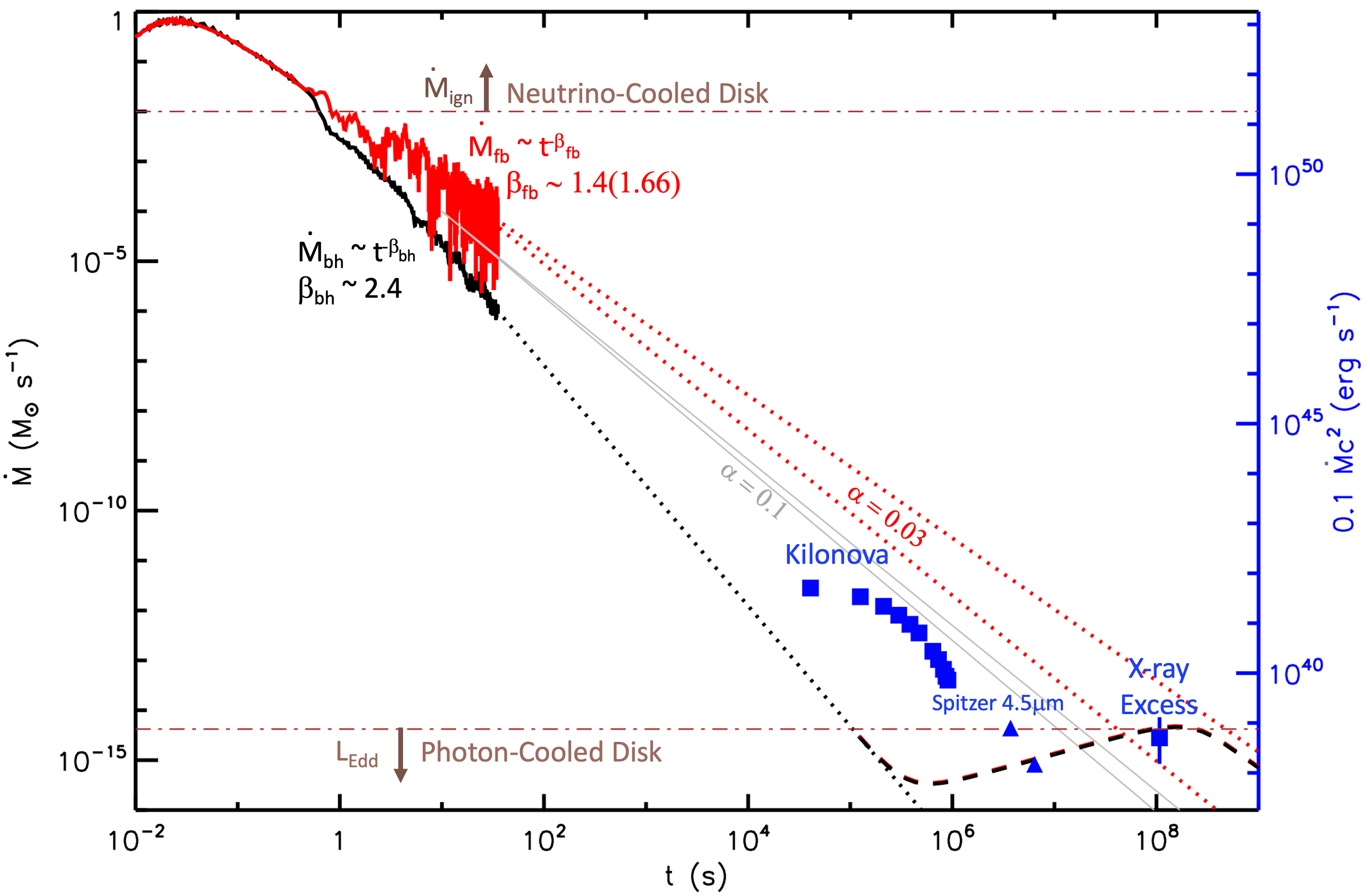}
    \caption{Connection between long-term accretion and X-ray excess in 
    GW170817. The black solid line shows the black hole accretion rate at the ISCO $\dot{M}_{\rm bh}$ for our fiducial $\alpha = 0.03$ calculation, while the red solid line shows the rate of decrease of gravitationally bound mass due to disk outflows $\dot{M}_{\rm fb}$(gray solid lines show the equivalent for the $\alpha = 0.1$ simulations).  The black and red dotted lines show the best-fit power-law extrapolations of $\dot{M}_{\rm bh}$ and $\dot{M}_{\rm fb}$, respectively, to later times (eqns.~\ref{eq:pl_bh} and \ref{eq:pl_fb}).  The right axis shows the corresponding accretion power $L_{\rm acc} \equiv 0.1 \dot{M}_{\rm bh} c^{2}$ as a proxy for maximum X-ray luminosity released through disk accretion onto the black hole.  We expect $\dot{M}_{\rm bh}$ and $\dot{M}_{\rm fb}$ to roughly track each other at early times (above the top brown dot-dashed line) when the disk is efficiently neutrino-cooled and at late times (below the brown dot-dashed line, $L_{\rm acc} < L_{\rm Edd}$) when the disk is photon-cooled.  Shown for comparison with a black dashed line is the inferred luminosity of the X-ray excess from GW170817 \citep{Hajela+21,Troja+21}.  Also shown for comparison with blue points is the kilonova luminosity, as estimated bolometrically across the optical-NIR bands during the first weeks \citep{Cowperthwaite+17} and as a lower-limit at later times based on {\it Spitzer} 4.5$\mu$m detections \citep{Kasliwal+19,Villar+18}.}
    \label{fig:fallback}
\end{figure*}

\section{Simulations}

We simulate the long-term evolution of the accretion disk from GW170817 assuming that a black hole formed promptly. The time-dependent evolution is carried out in axisymmetry with the hydrodynamic
code FLASH version 3.2 \citep{fryxell00,dubey2009}. The public version has been modified to include angular momentum transport with an
imposed shear stress, 
following the prescription of \citet{shakura1973} and \citet{stone1999}, 
neutrino emission and absorption using a leakage scheme for emission 
and an annular lightbulb for absorption \citep{Fernandez&Metzger13,MF14}, 
and a pseudo-Newtonian potential to model the gravity of the black hole \citep{artemova1996,FKMQ14}. The
equation of state is that of \citet{timmes2000}, modified such that 
neutrons, protons, and alpha particles are in nuclear 
statistical equilibrium for $T>5\times 10^9$\,K, and accounting for
changes in nuclear binding energy (nuclear dissociation and recombination).

The initial condition is a black hole of mass $M_{\rm bh} = 2.65M_\odot$ and dimensionless
spin $0.8$, as well as an equilibrium torus with initial mass $0.1M_\odot$,
consistent with values inferred for GW170817 \citep{LVSC2017a,Shibata2017}.
The torus has constant initial electron fraction $0.1$, 
entropy $8$\,k$_{\rm B}$ per baryon, 
and constant specific angular momentum. The domain outside the
torus is filled with a low-density ambient medium. The boundary
conditions are outflow in radius and reflecting in polar angle
at the symmetry axis. The radial range extends from a radius halfway 
between the ISCO and the horizon, out to a radius $10^5$ times larger.
The grid is discretized with 640 cells logarithmically spaced in radius and
with 112 cells uniformly spaced in $\cos\theta$.

We run two simulations using values of the viscosity
parameter $\alpha=\{0.03,0.1\}$, 
which bracket the late-time ($t>1$\,s) behavior
of the accretion rate in GRMHD simulations \citep{Fernandez+19}.
%The model with $\alpha=0.03$ is 
Both models are
evolved for $10,000$ orbits at the initial density
maximum of the torus ($50$\,km), corresponding to $\sim 35$\,s. 
%The model with $\alpha=0.1$ is currently at $30$\,s.
By this time, neutrino emission has decreased 
to negligible levels and
and nuclear dissociation/recombination has ceased, since the maximum 
temperature in the
disk is less than $5\times 10^9$\,K.  As summarized in Table \ref{t:models}, the masses ejected with positive energy\footnote{Requiring a positive Bernoulli
parameter instead of positive energy increases the ejecta for the $\alpha=0.03$
model to $0.02M_\odot$ while leaving the $\alpha=0.1$ ejecta 
nearly unchanged.}
 at a distance $10^9$\,cm from the BH by the $\alpha=\{0.03,0.1\}$ simulations are $\{1.7\times 10^{-2},3\times 10^{-2}\}M_{\odot}$, respectively, on the 
lower end of the range of $r$-process production inferred from the kilonova of GW170817 (e.g., \citealt{Villar+17}).

\section{Results}

\begin{table*}
\centering
\caption{Simulation parameters and results.}
\begin{tabular}{lcccccccccccc}
\hline
Model & $M_{\rm bh}$ & $M_{\rm d}$ & $\alpha$ & $t_{\rm max}$ & $M_{\rm ej}$ & $M_{\rm bnd}(t_{\rm max})$ & $\left.\frac{dM}{dE}\right|_{E = 0,t_{\rm max}}$ & $\dot{M}_{\rm bh,10}$ & $\beta_{\rm bh}$ & $\dot{M}_{\rm fb,10}$ & $\beta_{\rm fb}$ & $L_{\rm X}$\\
      & ($M_\odot$)  & ($M_\odot$) &          &    (s)    & ($M_{\odot}$) & ($M_{\odot}$) &  (g$^{2}$ erg$^{-1}$) &  ($M_\odot$\,s$^{-1}$) & & ($M_\odot$\,s$^{-1}$) & & (10$^{38}$ erg\,s$^{-1}$)\\
\hline
v03   &  2.65 & 0.10 &  0.03 & 35 & 1.7E-2 & 2.1E-3 & 1.1E+14 & 2.1E-5 & 2.41 & 4.0E-4 & 1.43 & 2$-$70 \\
v10   &  2.65 & 0.10 &  0.10 & 35 & 3.0E-2 & 4.9E-4 & 1.7E+13 & 1.0E-5 & 2.10 & 1.0E-4 & 1.72 & {\color{red} 0.1$-$0.5}
\end{tabular}
\label{t:models}
\begin{flushleft}
{\bf Note:} Columns from left to right show: model name, black hole mass,
initial disk mass, viscosity parameter, maximum simulation time, 
ejected mass with positive net energy, mass in the computational domain
with negative net energy at $t_{\rm max}$, 
$dM/dE$ around $E = 0$ and $t_{\rm max}$ (Fig.~\ref{f:dMde}),
normalization and slope of power-law fit to accretion rate at the
ISCO radius (eq.~\ref{eq:pl_bh}), normalization and slope of power law
fit to the rate of change of bound mass in the domain (eq.~\ref{eq:pl_fb}),
and X-ray luminosity at $t = 3.4$ years ($t\simeq 10^8$\,s), calculated from eq.~(\ref{eq:Lx}) adopting the normalization $\dot{M}_{\rm fb,10}$ at $t = 10$\,s, and for power-law decay indices ranging from $\beta = \beta_{\rm fb}$ to $\beta = 5/3$. 
\end{flushleft}
\end{table*}

\begin{figure*}
\includegraphics*[width=\textwidth]{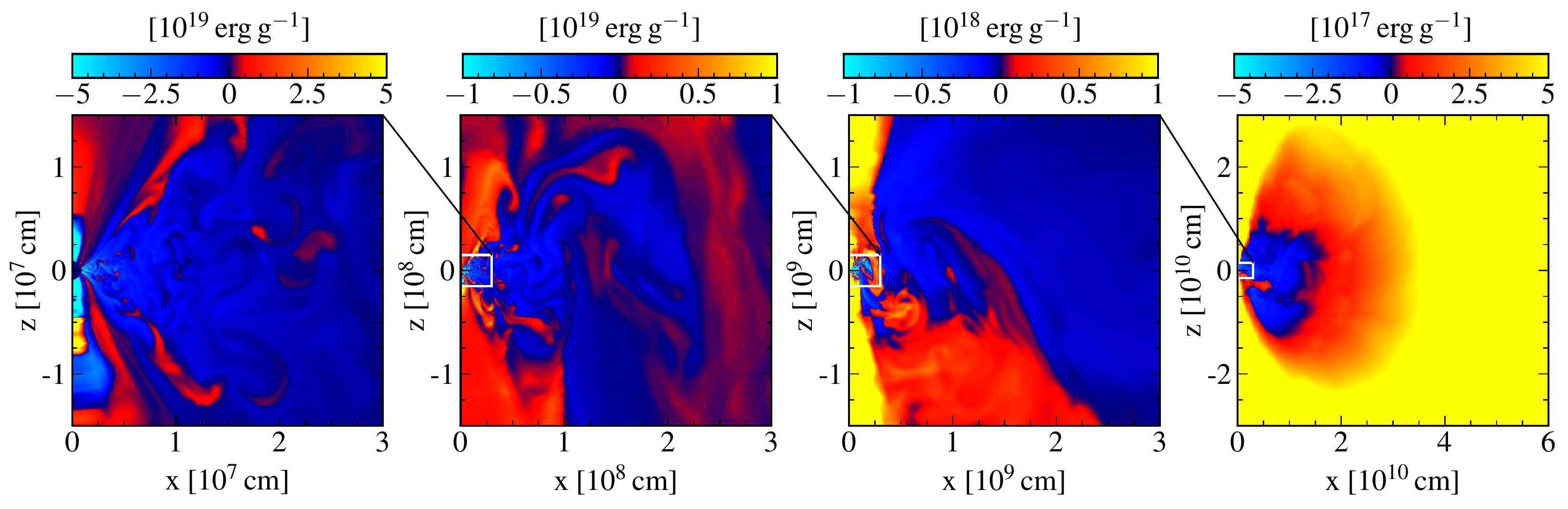}
\caption{Total specific energy (kinetic + internal + gravitational) at 
$t=35$\,s in the fiducial $\alpha=0.03$ model. Each of the three leftmost
panels are zoom-ins of the panels to their right, as shown.}
\label{f:etot_snapshots}
\end{figure*}

\begin{figure}
\includegraphics*[width=\columnwidth]{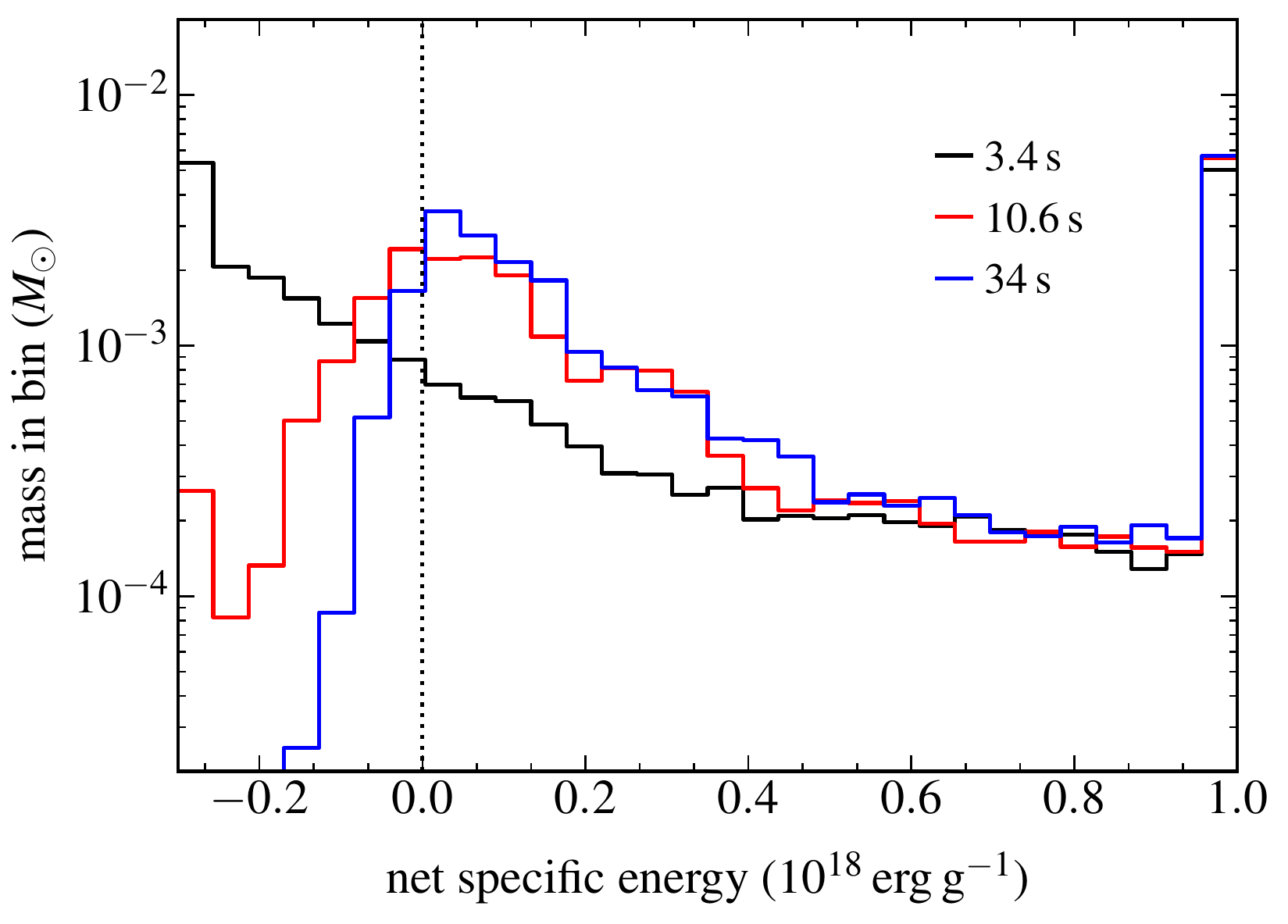}
\caption{Mass histogram of net specific energy (kinetic + internal + gravitational) for all matter in the computational domain for the fiducial $\alpha=0.03$ model at different times, as labeled. The edge bins contain all the material with energies beyond the range shown. The bin width is $\sim 4.3\times 10^{16}$\,erg\,g$^{-1}$, and the average of the bins around zero energy is $2.6\times 10^{-3}M_\odot$, thus  $dM/dE$ around zero energy is approximately $1.1\times 10^{14}$\,g$^2$\,erg$^{-1}$.}
\label{f:dMde}
\end{figure}

Figures \ref{fig:fallback}-\ref{f:dMde} summarize our results.  We focus on the $\alpha = 0.03$ model, insofar as it more closely matches the late-time 
accretion rate into the BH in GRMHD simulations \citep{Fernandez+19} and is the most promising in terms of generating a larger quantity of late-time fall-back accretion (we return to a discussion of the $\alpha = 0.1$ model at the end of the section).  In Fig.~\ref{fig:fallback}, a black solid line shows the accretion rate at the ISCO $\dot{M}_{\rm bh}$, while the red solid line shows the rate of decrease of gravitationally bound matter in the computational domain, $\dot{M}_{\rm fb}$.  The difference, $\dot{M}_{\rm out} = \dot{M}_{\rm fb}-\dot{M}_{\rm bh}$ represents mass unbound in outflows from the disk.  Dotted lines in Fig.~\ref{fig:fallback} show power-law extrapolations of $\dot{M}_{\rm bh}$ and $\dot{M}_{\rm fb}$ 
to later times than followed by our simulations, of the form (see Table \ref{t:models}).
\begin{eqnarray}
\label{eq:pl_bh}
\dot{M}_{\rm bh} & = &\dot{M}_{\rm bh,10}\left( \frac{t}{10\,\textrm{s}}\right)^{-\beta_{\rm bh}}\\
\label{eq:pl_fb}
\dot{M}_{\rm fb} & = &\dot{M}_{\rm fb,10}\left( \frac{t}{10\,\textrm{s}}\right)^{-\beta_{\rm fb}}.
\label{eq:fit}
\end{eqnarray}
where the fit parameters are obtained from $t=10$\, until
the end of each simulation. Note that the outflow
rate inferred from the decrease of the total bound mass 
in the computational domain computed here ($\dot{M}_{\rm fb}$) 
differs from the unbound mass flux at a fixed sampling radius, 
as typically reported in disk studies, since the latter decays
more steeply with time than the former as the bound disk
expands viscously past the sampling radius (e.g. Figure~\ref{f:etot_snapshots})
and outflows emitted at larger radii are missed.

The evolution of the torus follows that described in previous works (e.g., \citealt{Fernandez&Metzger13}).  At early times, the accretion rate is higher
than the characteristic value   
$\dot{M}_{\rm ign} \sim 10^{-2}M_{\odot}$ s$^{-1}$ 
(brown dot-dashed line in Fig.~\ref{fig:fallback}) above which 
the disk is sufficiently hot and dense to cool by neutrinos (e.g., \citealt{Narayan+01,Chen&Beloborodov07,De&Siegel21}), resulting in a high accretion efficiency ($\dot{M}_{\rm fb} \approx \dot{M}_{\rm bh}$; $\dot{M}_{\rm out} \ll \dot{M}_{\rm bh}$).  However, at later times as the accretion rate drops 
$\dot{M}_{\rm bh} < \dot{M}_{\rm ign}$, weak interactions become slow and freeze out, causing a transition to a radiatively inefficient state (e.g., \citealt{Metzger+08}). Once the maximum temperature in the disk decreases
below $5\times 10^9$\,K, internal energy changes due to nuclear dissociation
or recomination of alpha particles also stops.
At this point, viscous heating is no longer balanced by any significant 
cooling process and the disk becomes susceptible to outflows which carry much of its mass in unbound outflows, i.e. $\dot{M}_{\rm out} \approx \dot{M}_{\rm fb}$, suppressing accretion onto the black hole $\dot{M}_{\rm bh} \ll \dot{M}_{\rm fb}$ (e.g., \citealt{Blandford&Begelman99,Quataert&Gruzinov00}).  

Figure~\ref{f:etot_snapshots} shows a snapshot of the net specific energy $E$ at the end of the $\alpha = 0.03$ simulation ($t \approx 35$ s).  In the equatorial plane, gravitationally bound matter ($E < 0$) released by outflows during the early phases of the torus evolution, extends out to radii $r_{\rm out} \sim 10^{10}$ cm where $E \approx 0$.  The marginally bound matter at these large radii is not rotationally supported, but instead moves ballistically with a free-fall time $t_{\rm orb} \sim 2\pi(r_{\rm out}^{3}/GM)^{1/2}$ comparable to the system age $\sim t$.  Fresh matter can become unbound from the system ($E > 0$) only as a result of viscous heating, which occurs primarily in the rotationally supported disk on smaller scales $r_{\rm disk} \lesssim 3\times 10^{7}$ cm.  Pockets of matter which reach positive energy at $r \sim r_{\rm disk}$ expand outwards along the top and bottom boundary between the disk and polar funnel, where it mixes and joins a quasi-spherical unbound outflow at $r \gtrsim r_{\rm out}$ (see, e.g.,
Fig.~5 in \citealt{Fernandez+15}).

With this picture in mind, our interpretation for the evolution of the unbound mass, $\dot{M}_{\rm fb}$, is as follows.  Marginally bound debris, released during early phases in the disk evolution and extending out to radii $r_{\rm out}$ set by the condition $t_{\rm orb} \sim t$, is continuously falling back to join the inner rotationally-supported disk.  However, upon returning this material is promptly unbound by viscous heating due to the radiatively inefficient nature of the accretion flow and the low binding energy of the fall-back matter at late times (e.g., \citealt{Blandford&Begelman99,Li+13}), as suggested by \citet{Rossi&Begelman09}.  Thus, matter becomes unbound at a rate which follows the rate it falls back to the inner disk.

To check this interpretation, Figure~\ref{f:dMde} shows a mass histogram
of net specific energy $E$ at three different times in the $\alpha=0.03$
simulation. Of particular interest is the value of $dM/dE$ near 
$E \approx 0$, as this quantity controls the rate of mass fall-back to the disk of the marginally bound debris.  If mass falls back to the disk ballistically on the orbital time $t = t_{\rm orb} = 2\pi (a^{3}/GM_{\rm bh})^{1/2}$, where $a$ is the semi-major axis and $|E| = GM_{\rm bh}/2a$, then the fall-back rate is given by (e.g., \citealt{Rees88})
\be
\frac{dM}{dt} \approx 4\times 10^{-4}M_{\odot}{\rm s^{-1}}\,\, \left(\frac{dM/dE|_{E \approx 0}}{\rm 10^{14}\,g^{2}\,erg^{-1}}\right)\left(\frac{t}{10\,{\rm s}}\right)^{-5/3}
\label{eq:Mdotfb}
\ee
By the final two snapshots, the energy distribution is not evolving significantly around $E \approx 0$.  Using the value $dM/dE|_{E \approx 0} \approx 1.2\times 10^{14}$ g$^{2}$ erg$^{-1}$ inferred from the final snapshot, we find $dM/dt \approx 5\times 10^{-4}M_{\odot}$ at $t = 10$ s.  This is in excellent agreement with the rate at which matter is being unbound from the disk on the same timescale ($\dot{M}_{\rm fb,10}$; Table \ref{t:models}).  However, the best-fit power-law index for $\dot{M}_{\rm fb}$ of $\beta_{\rm fb} \approx 1.43$ (eq.~\ref{eq:fit}) is somewhat shallower than the value $\approx 5/3$ predicted by Eq.~(\ref{eq:Mdotfb}) for constant $dM/dE$.

Naive extrapolation of the black hole accretion rate (dotted black line in Fig.~\ref{fig:fallback}) to the present epoch $t \approx 3.4$\,yr would fall short of that required to explain the observed X-ray excess by many orders of magnitude, if one assumes
\be 
\label{eq:Lx}
L_X \approx \eta \dot{M}_{\rm bh}c^{2}, 
\ee
where radiative efficiency of $\eta \approx 0.1$ is the maximum value for a black hole of spin $0.8$ (\citealt{Bardeen+72}).  However, by the epoch $t \approx 3.4$\,yr, the rate of fall-back accretion has become sub-Eddington ({\it brown dot-dashed line} in Fig.~\ref{fig:fallback}), at which point the photons are no longer trapped in the flow and are able to diffuse out, radiatively cooling the disk (e.g., \citealt{Begelman79}).  Thus, as with the early neutrino-cooled epoch, we expect that $\dot{M}_{\rm bh}$ will again rise to match $\dot{M}_{\rm fb}$ starting a few years after the merger, in which case $L_{\rm X} \approx \eta \dot{M}_{\rm fb}c^{2}$ for $L_{\rm X} \lesssim L_{\rm Edd} \approx 8\times 10^{38}$ erg s$^{-1}$.  This transition is illustrated schematically with a dashed line in Fig.~\ref{fig:fallback}.  

Interpreted in this way, we see that the black hole accretion rate at $t \approx 3.4$ years may indeed be sufficient to explain the X-ray excess from GW170817, which we show for comparison with a blue square in Fig.~\ref{fig:fallback} \citep{Hajela+21,Troja+21}.  We include a larger error bar on the luminosity value than the formal one (by a factor of 2 in either direction), to account for uncertainties in (1) the bolometric correction of the disk emission into the {\it Chandra} bandpass; (2) residual absorption by the kilonova ejecta; (3) geometric beaming of the X-ray emission by the $\sim L_{\rm Edd}$ accretion flow.  

The lateness of the transition to a radiative efficient disk is also important to not over-producing the kilonova emission.  The kilonova ejecta will remain opaque to X-rays from the inner accretion disk for at least the first year of the explosion, only becoming optically thin around the present epoch \citep{Margalit+18,Hajela+21}.  Most of the accretion power that emerges during earlier epochs will be absorbed and reprocessed into optical/infrared emission (e.g., \citealt{Kisaka+16,Matsumoto+18}).  It is thus of equal importance that the maximal accretion power $\sim 0.1\dot{M}_{\rm bh}c^{2}$ remains below the observed luminosity of the kilonova emission.  The latter is shown in Fig.~\ref{fig:fallback} by blue squares where UVOIR data is available (e.g., \citealt{Cowperthwaite+17}), and as blue triangles where 4.5$\mu$m detections serve as lower-limits \citep{Kasliwal+19,Villar+18}.  Although the results are sensitive to how rapidly the flow becomes radiatively efficient, we cannot rule out that some or all of the infrared emission detected by {\it Spitzer} is powered by accretion instead of radioactivity.  

Although our $\alpha = 0.03$ model appears promising to explain the X-ray excess of GW170817, the $\alpha = 0.1$ model is less so (see Table \ref{t:models}).  The rate at which matter is unbound ($\sim$ falling back) in our $\alpha = 0.1$ simulations decays more steeply, $\beta_{\rm fb} \approx 1.7$ compared to the $\alpha = 0.03$ case, while its normalization (derived either directly from the rate at which fallback is unbound or from the energy distribution $dM/dE$ around $E \approx 0$), is a factor of $\approx 3-7$ times lower.  We surmise that the reason for
this difference is that a higher value of $\alpha$ leads to more vigorous heating of the debris which returns to the rotationally supported disk, thereby launching more powerful outflows which act to unbind material that would otherwise fall back.
This is consistent with the larger mass ejected by the model with $\alpha=0.1$
(Table~\ref{t:models}). We also find that at late-times the $\alpha=0.1$ model displays large amplitude (factor $\sim 10$) oscillations in the
accretion rate on timescales of $\sim 10$\,s, making our power-law fits to the
temporal dependence less reliable than in the $\alpha=0.03$ case.

An obvious caveat to our conclusions is that we have employed an 
$\alpha$-viscosity prescription in lieu of a self-consistent treatment of angular momentum transport via the magneto-rotational instability (MRI).  MHD simulations of the post-merger disk evolution which resolve the MRI find larger mass ejection than the $\alpha$-viscosity hydro simulations (e.g., \citealt{Siegel&Metzger17,Fernandez+19,Christie+19}).  
Mass outflow rates are usually measured at a fixed
extraction radius, but as pointed out earlier, over long timescales the
bound portion of the disk expands outward beyond any extraction radius
and outflow rates cut off steeply with time at that location. 
While the time-dependence
of outflow rates in MHD and hydrodynamic simulations is similar
when measured in this way \citep{Fernandez+19}, one cannot
assume that the time-dependence of the bound mass in the computational
domain will be the same owing to the additional mass ejection provided
by magnetic processes other than dissipation of MRI turbulence, which
are not present in hydrodynamic simulations.
On the other hand, accretion disk masses up to $\approx 0.2M_{\odot}$, twice as large as those adopted in our simulations, are compatible with numerical relativity merger simulations and the quantity of disk wind ejecta in GW170817 (e.g., \citealt{Shibata2017}).  Our simulations also neglect the impact of additional heating of the bound debris by the late-time radioactive decay of $r$-process nuclei, which can potentially impact the matter energy distribution and rate of late-time fall-back (e.g., \citealt{Metzger+10b,Desai+19,Ishizaki+21b}).

\section{Conclusions}

The surprisingly close neutron star merger GW170817 offers a unique opportunity to witness a single accretion flow evolve from an neutrino-cooled phase in the first second after the merger to a photon-cooled state a few years later, in between which is a long period of radiatively inefficient accretion.  Using long-term axisymmetric hydrodynamical simulations of the post-merger accretion disk system, we have taken preliminary steps to evaluate the feasibility of late-time accretion onto the newly formed black hole as a source of the recently discovered excess X-ray luminosity observed at 3.4 years after the merger.  Taking the rate of decline of bound gaseous mass as a proxy for the rate at which matter is returning to the disk at late times (and being unbound in disk winds), we find that a power-law extrapolation of $\dot{M}_{\rm fb}$ in our $\alpha = 0.03$ simulation, yields a late-time accretion rate onto the black hole which is in principle sufficient to explain the excess.  
%However, the estimated fall-back rate in our $\alpha = 0.1$ simulation appears to fall significantly short, demonstrating the sensitivity of our finding to the accretion physics.   

One prediction of the accretion powered origin is that the X-ray excess should, possibly after a brief brightening period (if the ejecta is still marginally opaque to X-rays at the present epoch), begin to decay, roughly as $L_{\rm X} \propto t^{-\beta}$ where $\beta \approx 5/3$ \citep{Hajela+21}.  This contrasts with alternative kilonova afterglow model, for which $L_{\rm X}$ is predicted to decay slower, or even rise, in time (e.g., \citealt{Kathirgamaraju+19,Nedora+21}), and should eventually be accompanied by radio emission in excess of the ordinary afterglow contribution.

%We furthermore show how suppression of the black hole accretion rate at earlier epochs, due to inefficient cooling, is key to not over-contributing to the observed kilonova through reprocessing (which we find must still be dominated by energy input from radioactive decay, except possibly at the final {\it Spitzer} epoch, where accretion power may contribute).  On the other hand, an extrapolation of our $\alpha = 0.1$ simulation data for $\dot{M}_{\rm fb}$ sizably underpredicts the required X-ray accretion luminosity.

\acknowledgements

BDM acknowledges support from NSF (grant AST-2002577) and NASA (grants NNX17AK43G, 80NSSC20K0909).
RF acknowledges support from the Natural Sciences and Engineering Research
Council (NSERC) of Canada through Discovery Grant RGPIN-2017-04286, and by the
Faculty of Science at the University of Alberta.
The software used in this work
was in part developed by the U.S. Department of Energy (DOE)
NNSA-ASC OASCR Flash Center at the University of Chicago.
Data visualization was done in part using {\tt VisIt} \citep{VisIt}, 
which is supported by DOE with funding from the Advanced Simulation 
and Computing Program and the Scientific Discovery through Advanced Computing Program. This research was enabled in part by support provided by WestGrid
(www.westgrid.ca), the Shared Hierarchical Academic Research Computing Network
(SHARCNET, www.sharcnet.ca), Calcul Qu\'ebec (www.calculquebec.ca), and Compute
Canada (www.computecanada.ca).
This research also used resources of the U.S. 
National Energy Research Scientific Computing
Center (NERSC), which is supported by the DOE Office of Science 
under Contract No. DE-AC02-05CH11231 (repository m2058).

%\bibliographystyle{mn2e}
%\bibliographystyle{aasjournal}
%\bibliography{refs,rodrigo}

\end{document}